\documentclass[12pt,a4paper]{article}
\usepackage[utf8]{inputenc}
\usepackage{amsmath,amsfonts,amssymb}
\usepackage{graphicx}

\usepackage[backend=bibtex,giveninits=true,sorting=nyt]{biblatex}
\newcommand{\deriv}[1]{\frac{\textrm{d}#1}{\textrm{d}t}}

\title{A data driven analysis and forecast of an SEIARD epidemic model for COVID-19 in Mexico}

\author{
    Ugo Avila-Ponce de Le\'on$^1$,
    \'Angel G. C. P\'erez$^2$,
    Eric Avila-Vales$^2$,
}

\date{}

\begin{document}

\maketitle

\begin{center}
    {\small
        $^1$
        Programa de Doctorado en Ciencias Biol\'ogicas, Universidad Nacional Aut\'onoma de M\'exico, Mexico City, Mexico\\
        
        $^2$
        Facultad de Matem\'aticas, Universidad Aut\'onoma de Yucat\'an, Anillo Perif\'erico Norte, Tablaje Catastral 13615, C.P. 97119, M\'erida, Yucat\'an, Mexico
    }
\end{center}
\bigskip

\begin{abstract}
    We propose an SEIARD mathematical model to investigate the current outbreak of coronavirus disease (COVID-19) in Mexico. We conduct a detailed analysis of this model and demonstrate its application using publicly reported data. We calculate the basic reproduction number ($R_0$) via the next-generation matrix method, and we estimate the per day infection, death and recovery rates. We calibrate the parameters of the SEIARD model to the reported data by minimizing the sum of squared errors and attempt to forecast the evolution of the outbreak until June 2020. Our results estimate that the peak of the epidemic in Mexico will be around May 2, 2020. Our model incorporates the importance of considering the aysmptomatic infected individuals, because they represent the majority of the infected population (with symptoms or not) and they could play a huge role in spreading the virus without any knowledge.
\end{abstract}

\section{Introduction}

The COVID-19 pandemic originated in Wuhan, China in December 2019. Since then, the number of cases has accelerated in China and subsequently all over the world. The causative agent is a new betacoronavirus related to the Middle East Respiratory Syndrome virus (MERS-CoV) and the Severe Acute Respiratory Syndrome virus (SARS-CoV).

On January 30, the World Health Organization (WHO) formally declared the outbreak of novel coronavirus a Global Public Health Emergency of International Concern. In \cite{cruzpacheco2020dispersion}, \citeauthor{cruzpacheco2020dispersion} estimated the arrival of the infectious outbreak to Mexico between March 20 and March 30, 2020. Other models for predicting the evolution of COVID-19 outbreak in Mexico have been proposed in \cite{vivancolira2020predicting,alvarez2020modeling,acunazegarra2020sars}.

Compartmental models have been used for studying the spread of the COVID-19 pandemic in several countries, such as China \cite{lin2020conceptual,khrapov2020mathematical,fanelli2020analysis,caccavo2020chinese}, Italy \cite{fanelli2020analysis,caccavo2020chinese}, India \cite{singh2020age} and Brazil \cite{bastos2020modeling}.

\section{Mathematical model}

In this work, we will use a compartmental differential equation model for the spread of COVID-19 in Mexico. The model monitors the dynamics of six subpopulations, which are: susceptible ($S(t)$), exposed ($E(t)$), infected ($I(t)$), asymptomatic ($A(t)$), recovered ($R(t)$) and dead ($D(t)$).

The model simulations will be carried out with the following assumptions:
\begin{itemize}
    \item[(a)] Individuals of 12 years old and higher are susceptible to the virus.
    
    \item[(b)] The susceptible and infected individuals are homogeneous in the population.
    
    \item[(c)] At first, no interventions were applied to stop the spread of COVID-19.
    
    \item[(d)] The population is constant; no births are allowed, and we only take into account the fatalities associated to COVID-19.
\end{itemize}

\begin{figure}
    \centering
    \includegraphics[width=0.4\linewidth]{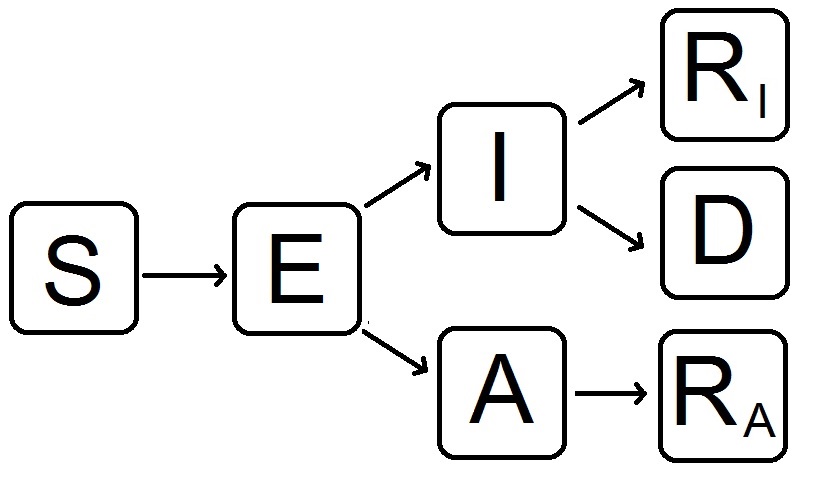}
    \caption{Flow diagram of our mathematical model to evaluate the behavior of the spread of nCoV-2019 in Mexico. $S$: susceptible, $E$: exposed, $I$: infected, $A$: infected but without symptoms (asymptomatic), $R_I$: recovered from symptomatic infection, $R_A$: recovered from asymptomatic infection, $D$: dead.}
    \label{fig:diagram}
\end{figure}

Figure \ref{fig:diagram} shows a diagram of the flow through the compartmental subpopulations.

\textbf{Susceptible population $S(t)$}: This subpopulation will remain constant because recruiting individuals is not allowed in our model. The susceptible population will decrease after an infection, an acquired characteristic due to the interaction with an infected person or asymptomatic one. The transmission coefficients will be $\beta I$ and $\beta A$. The rate of change of the susceptible population is expressed in the following equation:
\begin{equation}
\deriv{S} = -\beta S\left(\frac{I+A}{N-D}\right).
\end{equation}

\textbf{Exposed population $E(t)$}: This subpopulation consists of individuals that are infected but cannot infect others. The population decreases at a rate $w$ to become infected or asymptomatic. Consequently,
\begin{equation}
\deriv{E} = \beta S\left(\frac{I+A}{N-D}\right) - wE = \beta S\left(\frac{I+A}{N-D}\right) - pwE - (1-p)wE.
\end{equation}

\textbf{Infected population $I(t)$}: Infected (symptomatic) individuals are generated at a proportion $p$ from the exposed class. They recofer at a rate $\gamma$ and die at a rate $\delta$. This is the only population that acknowledges death. Thus,
\begin{equation}
\deriv{I} = pwE - (\delta+\gamma)I.
\end{equation}

\textbf{Asymptomatic population $A(t)$}: This population is considered an infected population, but the individuals do not develop the common symptoms of COVID-19. Asymptomatic individuals are important to model because they have the ability to spread the virus without knowing; they are produced at a rate $1-p$ and recover at a rate $\gamma$. Consequently,
\begin{equation}
\deriv{A} = (1-p)wE - \gamma A.
\end{equation}

\textbf{Recovered populations $R_I(t)$ and $R_A(t)$}: All individuals infected with symptoms or not will recover at a rate $\gamma$. We subdivide the recovered population in two compartments: individuals who recover after having symptoms ($R_I$) and individuals who recover from asymptomatic infection ($R_A$). Hence
\begin{equation}
\deriv{R_I} = \gamma I,
\qquad\deriv{R_A} = \gamma A.
\end{equation}

\textbf{Dead population $D(t)$}: Infected individuals with symptoms die at a rate $\delta$, that is,
\begin{equation}
\deriv{D} = \delta I.
\end{equation}

Hence, the system of differential equations that will model the dynamics of coronavirus spread in Mexico is:
\begin{equation}
\begin{aligned}
	\deriv{S}   & = -\beta S\left(\frac{I+A}{N-D}\right),     \\
	\deriv{E}   & = \beta S\left(\frac{I+A}{N-D}\right) - wE, \\
	\deriv{I}   & = pwE - (\delta+\gamma)I,                   \\
	\deriv{A}   & = (1-p)wE - \gamma A,                       \\
	\deriv{R_I} & = \gamma I,                                 \\
	\deriv{R_A} & = \gamma A,                                 \\
	\deriv{D}   & = \delta I.
\end{aligned}
\end{equation}

We also observe that $N:=S+E+I+A+R_I+R_A+D$ is constant, where $N$ is the size of the population modeled.

The rest of this work is organized as follows. In Section \ref{sec:basic}, we calculate the basic reproduction number of the model. We present in Section \ref{sec:implem} the assumptions and procedure we used for the implementation of our model, and we describe in Section \ref{sec:resul} the results obtained when performing simulations with the best fit parameters.

\section{Basic reproduction number}\label{sec:basic}

There exists a disease-free equilibrium, which is given by $S=N$, $E=I=A=R_I=R_A=D=0$, and we will denote it by $x_0$. We calculate the basic reproduction number $R_0$ based on this steady state. We use the next-generation matrix method proposed by Diekmann et al. \cite{diekmann2010construction}. To find $R_0$, we must solve the equation $R_0=\rho(FV^{-1})$, where $F$ and $V$ are the derivatives of the new infections matrix $\mathcal{F}$ and the transition matrix $\mathcal{V}$, respectively, evaluated at the disease-free equilibrium. Then
\[\mathcal{F}=
\begin{bmatrix}
	\beta S\left(\dfrac{I+A}{N-D}\right) \\
	0                                   \\
	0
\end{bmatrix}\]
The derivative of $\mathcal{F}$ at $x_0$ is:
\[F =
\begin{bmatrix}
	0 & \beta & \beta \\
	0 & 0     & 0     \\
	0 & 0     & 0
\end{bmatrix}.\]

The transition matrix is
\[\mathcal{V} =
\begin{bmatrix}
	wE                      \\
	-pwE + (\delta+\gamma)I \\
	-(1-p)wE + \gamma A
\end{bmatrix}.\]
The derivative of $\mathcal{V}$ at $x_0$ is
\[V =
\begin{bmatrix}
	w       & 0             & 0      \\
	-pw     & \delta+\gamma & 0      \\
	-(1-p)w & 0             & \gamma
\end{bmatrix}.\]
The inverse of $V$ is
\[V^{-1} =
\begin{bmatrix}
	\dfrac1w                 & 0                        & 0                 \\
	\dfrac{p}{\delta+\gamma} & \dfrac{1}{\delta+\gamma} & 0                 \\
	\dfrac{1-p}{\gamma}      & 0                        & \dfrac{1}{\gamma}
\end{bmatrix}.\]
Then
\[FV^{-1} =
\begin{bmatrix}
	\dfrac{\beta p}{\delta+\gamma} + \dfrac{\beta(1-p)}{\gamma} & \dfrac{\beta}{\delta+\gamma} & \dfrac{\beta}{\gamma} \\
	0                                                           & 0                            & 0                     \\
	0                                                           & 0                            & 0
\end{bmatrix}.\]

We need to find the eigenvalues of $FV^{-1}$, which are $\lambda_1 = \frac{\beta p}{\delta+\gamma} + \frac{\beta(1-p)}{\gamma}$, $\lambda_2 = 0$ and $\lambda_3 = 0$. Then, the basic reproduction number is given by the dominant eigenvalue, that is,
\begin{equation}\label{R0}
R_0 = \frac{\beta p}{\delta+\gamma} + \frac{\beta(1-p)}{\gamma}.
\end{equation}

\section{Implementation}\label{sec:implem}

To describe the evolution of the epidemic in Mexico taking into account the social distancing measures taken by the government, we will assume that the infection rate, recovery rate and death rate are time-dependent functions, similar to those used in \cite{caccavo2020chinese}.

To model the effect of epidemic control measures, which cause the number of contacts per person per unit time to decrease as the epidemic progresses, we describe the infection rate by the function
\begin{equation*}
\beta(t) = \beta_0\exp\left(-\frac{t}{\tau_\beta}\right) + \beta_1,
\end{equation*}
where $\beta_0+\beta_1$ is the initial infection rate. This rate decreases exponentially to the value $\beta_1$ with a characteristic time of decrease $\tau_\beta$.

The time of recovery for patients may also vary with time due to the medical staff improving their therapeutic procedures. Hence, we will assume that the recovery rate is modeled by the function
\begin{equation*}
\gamma(t) = \gamma_0 + \frac{\gamma_1}{1+\exp(-t+\tau_\gamma)},
\end{equation*}
where $\gamma_0$ is the recovery rate at time zero, and $\gamma_0 + \gamma_1$ is the recovery rate at a later time, which is reached after $\tau_\gamma$ days of adaptation.

Lastly, the death rate may decrease with time due to the adaptation of the pathogen or the development of more advanced treatments. Hence, we can model this with the function
\begin{equation*}
\delta(t) = \delta_0\exp\left(-\frac{t}{\tau_\delta}\right) + \delta_1,
\end{equation*}
where $\delta_0+\delta_1$ is the initial death rate, which decreases to the value $\delta_1$ with a characteristic time $\tau_\delta$.

If we replace the constant parameters $\beta$, $\delta$ and $\gamma$ in equation \eqref{R0} with the aforementioned time-dependent functions, we can define
\begin{equation}
R_d(t) = \frac{\beta(t)p}{\delta(t)+\gamma(t)} + \frac{\beta(t)(1-p)}{\gamma(t)}
\end{equation}
as the effective daily reproduction ratio, which measures the number of new infections produced by a single infected individual per day, taking into account the evolving public health interventions and available resources \cite{tang2020updated}.

The set of differential equations was solved using Matlab 2016b with the ode45 solver, which is based on an explicit Runge-Kutta (4,5) formula. Our model was calibrated using the cases of COVID-19 in Mexico. The data were collected in the period since the first reported case of COVID-19 in Mexico (February 28) until April 14 from the open source repository of Johns Hopkins University \cite{johnshopkins2019}.

The optimization of parameters to describe the outbreak of COVID-19 in Mexico were fitted by minimizing the Sum of Squared Errors (SSE), in such a way that the solutions for $D(t)$ and $R_I(t)$ obtained by the model approximate the reported values for deaths and recovered cases, respectively, while the sum $I(t) + R_I(t) + D(t)$ approximates the cumulative number of infected cases with symptoms. Since the Mexican government does not keep a record of the number of asymptomatic cases, we assume that the asymptomatic infected population is about nine times larger than the population with symptoms, based on government estimations.

We applied three searches to minimize the SSE function: a gradient-based method, a gradient-free algorithm, and finally, a gradient-based method. This method was necessary to obtain the local minimum. We adapted the code from Caccavo \cite{caccavo2020chinese} for our mathematical model.

\section{Results}\label{sec:resul}

\begin{figure}
    \centering
    \includegraphics[width=0.85\linewidth]{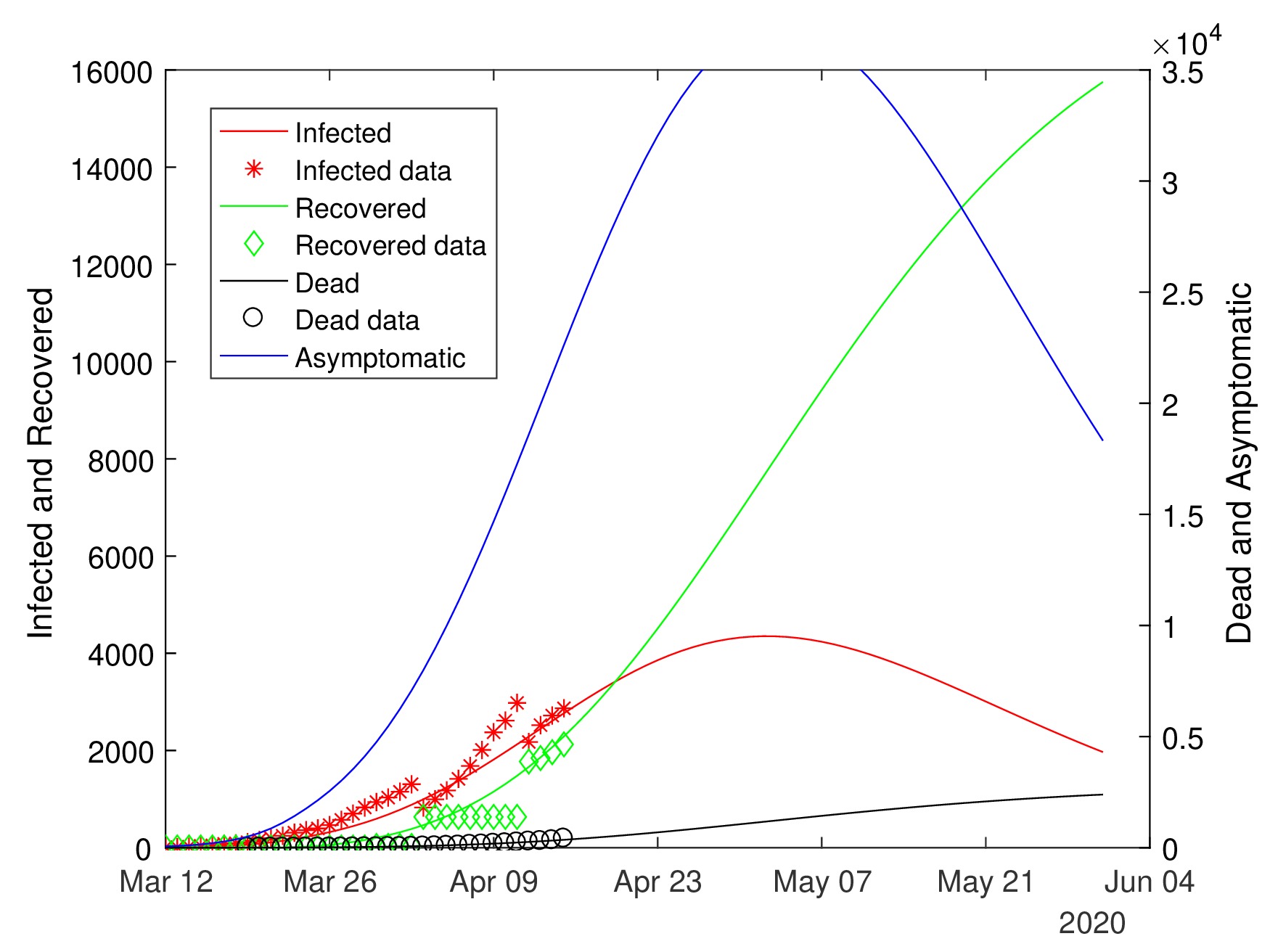}
    \caption{Graphs for the spread of COVID-19 in Mexico. Red dots represent the data for the infected. Green diamonds represent the data for the recovered individuals, and black circles are associated with the number of fatalities from the data. Solid lines with the same connotation in color represent the simulations of our model. The blue line denotes the estimated number of asymptomatic infections.}
    \label{fig:graphs}
\end{figure}

The predicted evolution of the outbreak for COVID-19 in Mexico can be seen in Figure \ref{fig:graphs}. The parameters of the mathematical model were fitted with the experimental data provided by a daily update from the Mexican Ministry of Health. By adjusting the data from the period from March 12, 2020 to April 14, 2020, we simulated the daily new COVID-19 cases in Mexico until June 4, 2020. The peak of the infection modeled will be around the first week of May, with 4400 infected individuals with the known developed symptoms and roughly 37\,000 infected individuals that will not develop any kind of symptoms.

\begin{figure}
    \centering
    \includegraphics[width=0.85\linewidth]{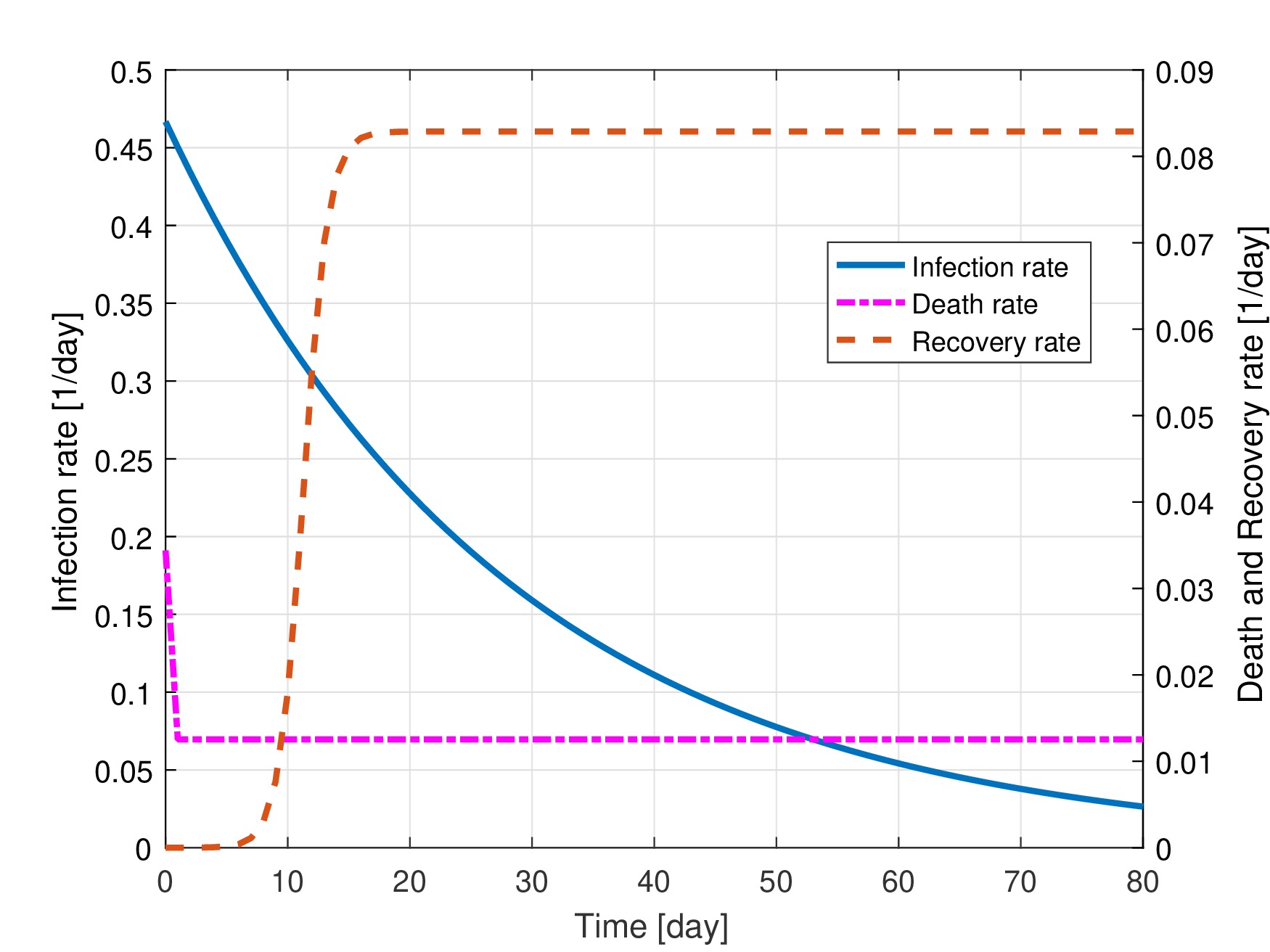}
    \caption{Best fit values of the infection, recovery and death rates as functions of time.}
    \label{fig:functions}
\end{figure}

\begin{figure}
    \centering
    \includegraphics[width=0.85\linewidth]{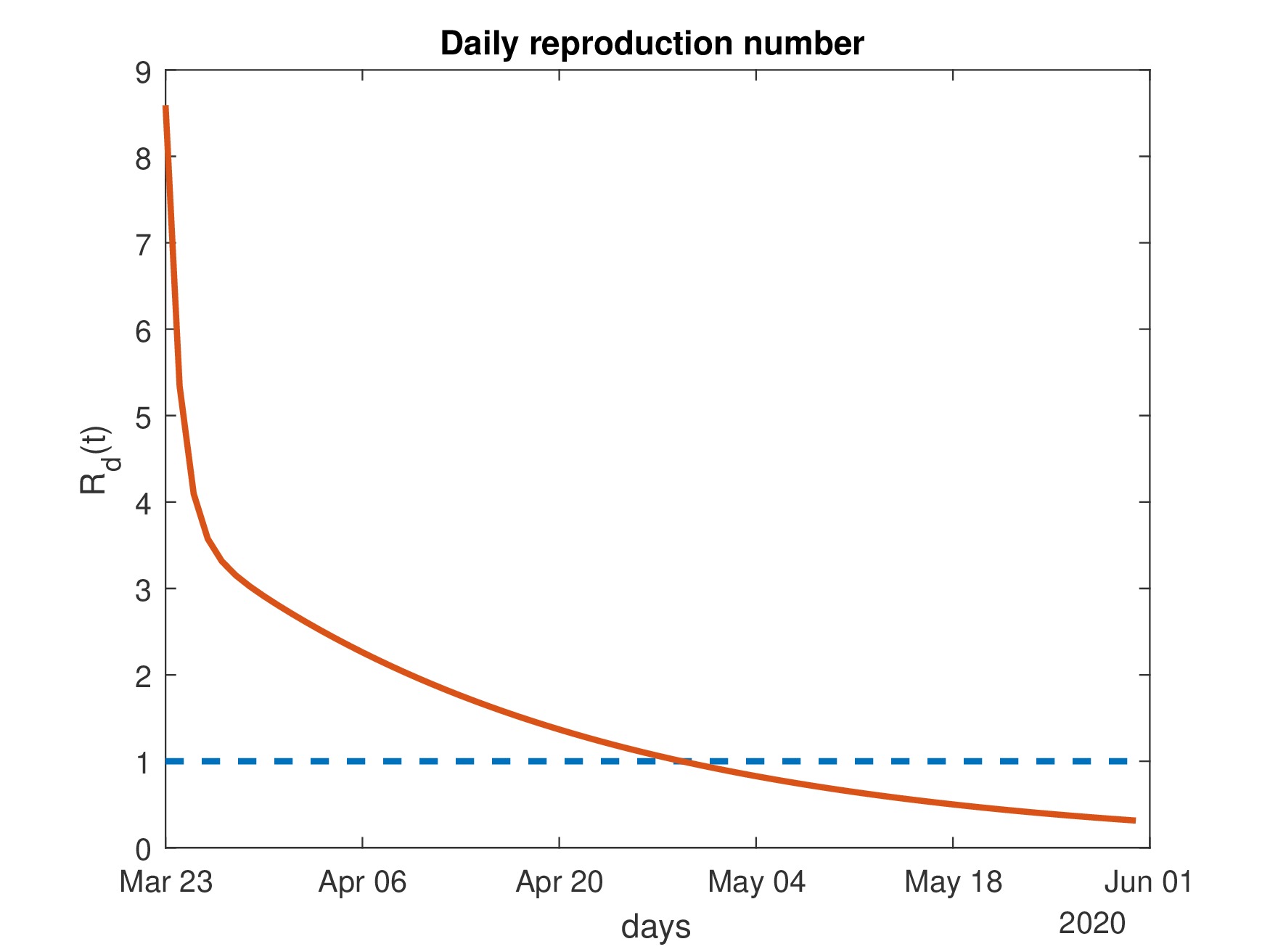}
    \caption{Variation of the effective daily reproduction number through time.}
    \label{fig:repnumber}
\end{figure}

The values of the best fit parameters are given in Table \ref{table1}. Figure \ref{fig:functions} shows the variation of the infection rate $\beta(t)$, recovery rate $\gamma(t)$ and death rate $\delta(t)$ with respect to time. Using these values for the parameters, we can calculate the effective daily reproductive ratio $R_d(t)$ for each day (see Figure \ref{fig:repnumber}). Our simulation shows that $R_d(t)$ will become less than 1 around April 29, 2020.

\begin{table}
    \caption{Model parameters obtained from the best fit optimization.}
    \label{table1}
    \centering
    \begin{tabular}{|c|c|c|}
    	\hline
    	  Parameter   &  Value  & Unit  \\ \hline\hline
    	  $\beta_0$   & 0.4668  & 1/day \\
    	  $\beta_1$   & 0.0100  & 1/day \\
    	$\tau_\beta$  & 27.8602 &  day  \\
    	 $\gamma_0$   & 0.0000  & 1/day \\
    	 $\gamma_1$   & 0.0829  & 1/day \\
    	$\tau_\gamma$ & 11.2885 &  day  \\
    	 $\delta_0$   & 0.0219  & 1/day \\
    	 $\delta_1$   & 0.0125  & 1/day \\
    	$\tau_\delta$ & 0.0003 &  day  \\
    	     $w$      & 0.9000  & 1/day \\
    	     $p$      & 0.1163  &  --   \\ \hline
    \end{tabular}
\end{table}

\begin{figure}
    \centering
    \includegraphics[width=0.85\linewidth]{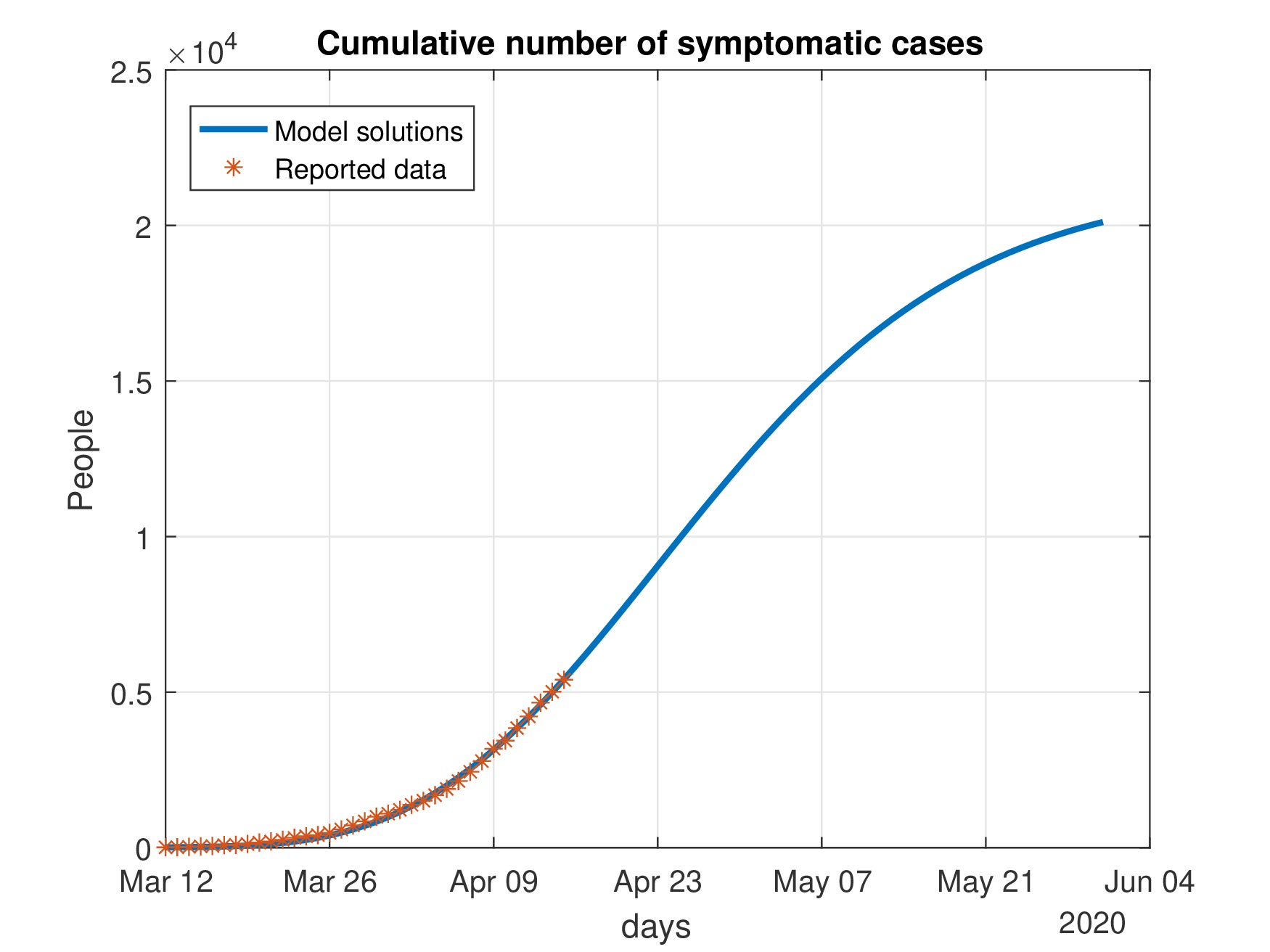}
    \caption{Cumulative number of symptomatic infected people ($I(t) + R_I(t) + D(t)$) predicted by the model and reported number of infected cases.}
    \label{fig:c}
\end{figure}

\begin{figure}
    \centering
    \includegraphics[width=0.85\linewidth]{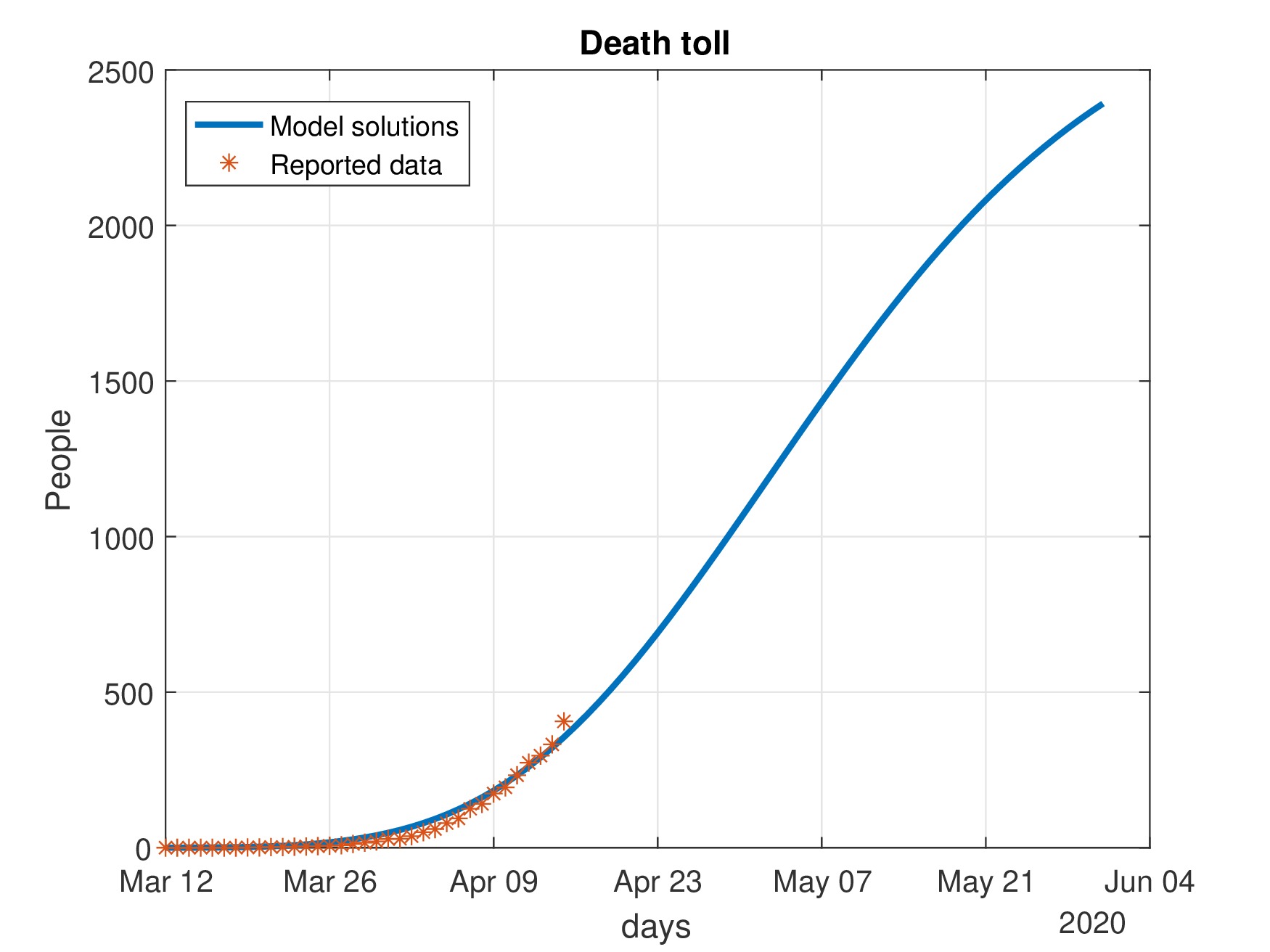}
    \caption{Death toll ($D(t)$) predicted by the model and reported number of deaths.}
    \label{fig:d}
\end{figure}

\begin{figure}
    \centering
    \includegraphics[width=0.85\linewidth]{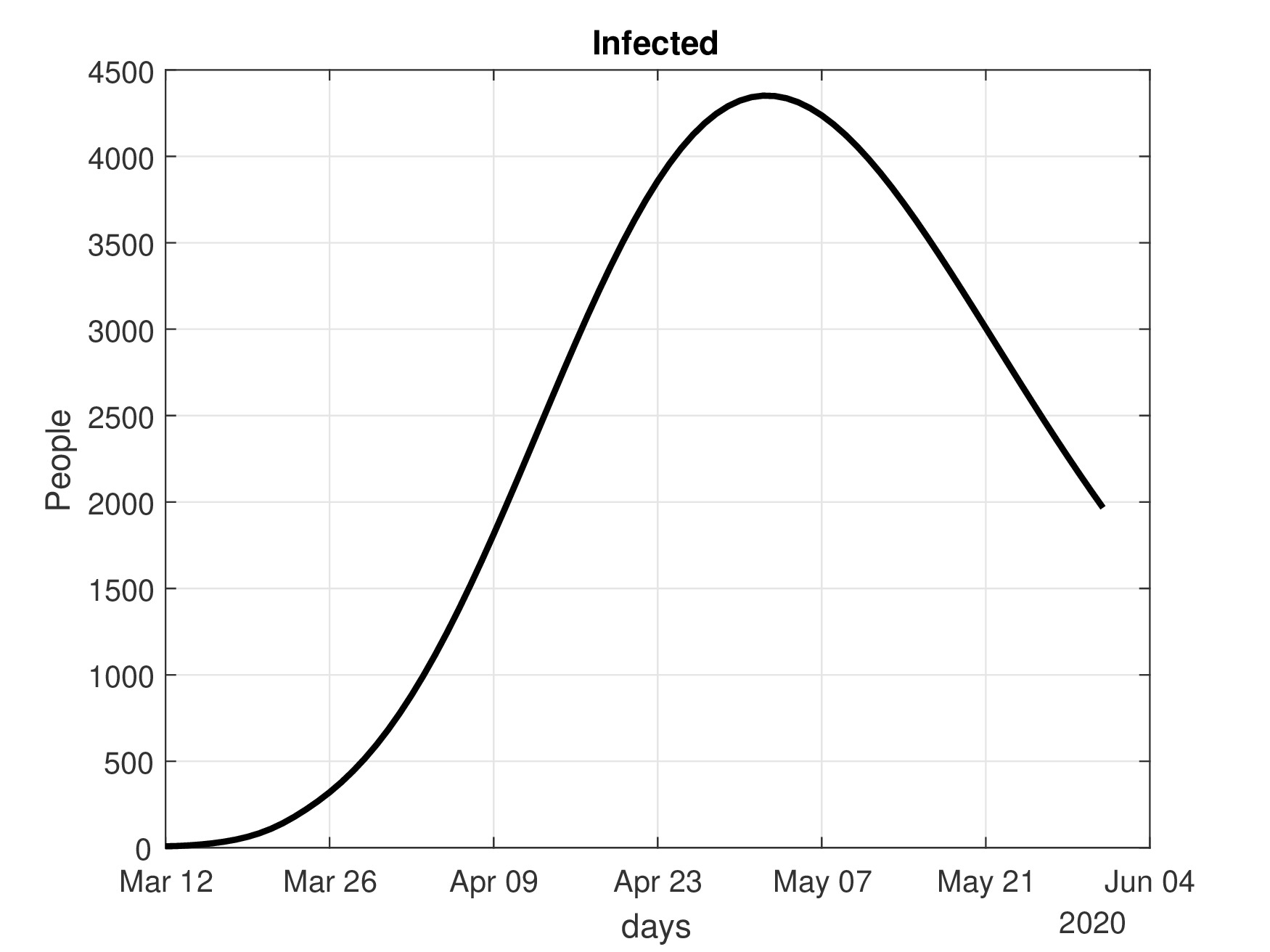}
    \caption{Number of infected cases ($I(t)$) predicted by the model.}
    \label{fig:i}
\end{figure}

\begin{figure}
    \centering
    \includegraphics[width=0.85\linewidth]{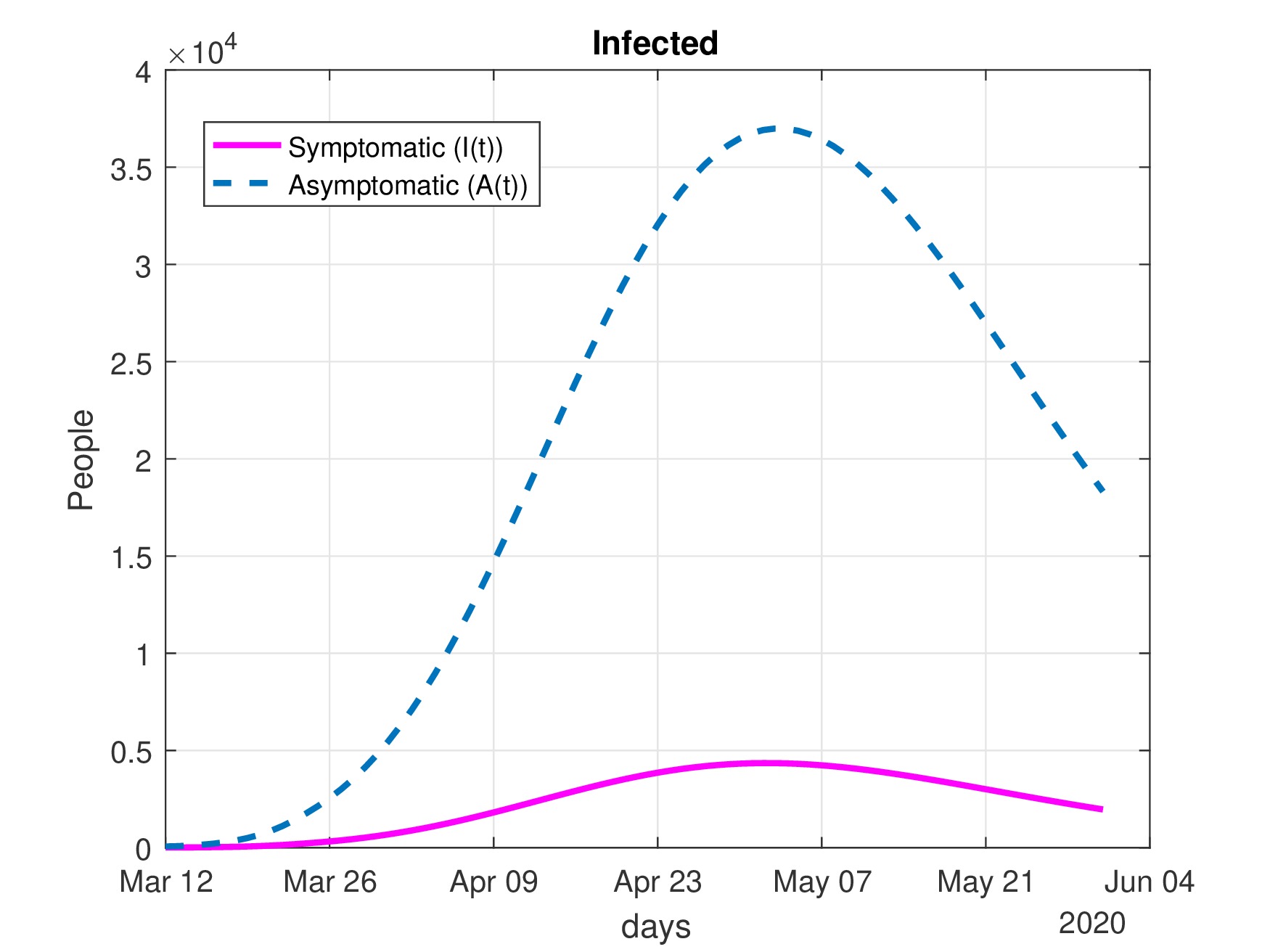}
    \caption{Number of infected cases ($I(t)$) and asymptomatic cases ($A(t)$) predicted by the model.}
    \label{fig:ia}
\end{figure}

In Figures \ref{fig:c}--\ref{fig:ia}, we carried out the simulation with the best fit parameters. We show a comparison of the cumulative number of infections (Figure \ref{fig:c}) and deaths (Figure \ref{fig:d}) with the reported data. We also plot the number of active symptomatic infections and asymptomatic infections in Figures \ref{fig:i} and \ref{fig:ia}.

As we can see, the population of asymptomatic individuals represents roughly 90\% of all possible infections. The behavior of the asymptomatic population is important because they have the capacity to spread the virus without developing any symptoms, and public health interventions should focus on these individuals. By this date, Mexico is located in phase 2 (community transmission) and has declared a national emergency and asked the citizenship to stay in their houses. By declaring self-quarantine to all individuals, we could separate the healthy population from the asymptomatic population and prevent the spread of the virus more rapidly. This action will only gain time in the number of infections and help avoiding the saturation of hospitals. If we control the population of infected individuals, we will prevent the death of many individuals, and the hospitals will have sufficient supplies to mitigate the severity of symptoms in patients with any type of chronic degenerative disease.

It is of great importance to explain some of the parameters estimated in the optimization section. $\tau_\beta=27.86$ represents the number of contacts per person per time. Even though at this time this value is high, by declaring social distancing measures the Mexican government intends to decrease the value of this parameter. We will view this decrease once we reach the peak of infection. $\tau_\gamma$ gives us information of the average recovery rate between the first day of the appearance of symptoms and death, which is $\tau_\gamma=11.28$. This means that some infected individuals can develop more complicated symptoms, eventually they will take much more time to recover than others. As well, it represents the average time of supplies a patient will cost if they develop chronic symptoms like pneumonia or other pulmonary complications.

\printbibliography

\end{document}